\documentclass[aip,apl,amsmath,amssymb,reprint]{revtex4-2}

\usepackage{graphicx}
\usepackage{dcolumn}
\usepackage{bm}

\usepackage[T1]{fontenc}
\usepackage{mathptmx}
\usepackage{etoolbox}
\usepackage{hyperref,xcolor}
\makeatletter
\def\@email#1#2{%
	\endgroup
	\patchcmd{\titleblock@produce}
	{\frontmatter@RRAPformat}
	{\frontmatter@RRAPformat{\produce@RRAP{*#1\href{mailto:#2}{#2}}}\frontmatter@RRAPformat}
	{}{}
}%
\makeatother
\def \bm#1{{\bf #1}}

\begin{document}

	\title[]
	{Jet-driven viscous locomotion of confined thermoresponsive microgels}
	\author{Ivan Tanasijevi\'c}
	
	\affiliation{Department of Applied Mathematics and Theoretical Physics, University of Cambridge, Wilberforce Road, Cambridge CB3 0WA, United Kingdom}
	
	\author{Oliver Jung}
	
	\affiliation{DWI Leibniz-Institute for Interactive Materials, RWTH Aachen University, Forckenbeckstr.~50, D-52056 Aachen,~Germany}
	
	\author{Lyndon Koens}
	
	\affiliation{Macquarie University, Macquarie Park, NSW 2113, Sydney, Australia}
	
	\author{Ahmed Mourran}
	
	\affiliation{DWI Leibniz-Institute for Interactive Materials, RWTH Aachen University, Forckenbeckstr.~50, D-52056 Aachen,~Germany}
	
	\author{Eric Lauga*}
	
	\email{e.lauga@damtp.cam.ac.uk}
	
	\affiliation{Department of Applied Mathematics and Theoretical Physics, University of Cambridge, Wilberforce Road, Cambridge CB3 0WA, United Kingdom}

	\date{\today}
	
	\begin{abstract}
		We consider the dynamics of micro-sized, asymmetrically-coated thermoresponsive hydrogel ribbons (microgels) under periodic heating and cooling in the confined space between two planar surfaces. As the result of the temperature changes, the volume and thus the shape of the slender microgel change, which lead to repeated cycles of bending and elastic relaxation, and to net locomotion. 		Small devices designed for biomimetic locomotion need to exploit flows that are not symmetric in time (non-reciprocal) to escape the constraints of the scallop theorem and undergo net motion.  Unlike other biological slender swimmers, the non-reciprocal bending of the gel centreline is not sufficient here to explain for the overall swimming motion. We show instead that the swimming of the gel results from the flux of water periodically emanating from (or entering) the gel itself due to its shrinking (or swelling). The associated flows induce viscous stresses that lead to a net propulsive force on the gel.
		We derive a theoretical model for this hypothesis of jet-driven propulsion, which leads to excellent agreement with our experiments.
		
	\end{abstract}
	
	\maketitle
	
	Self-propulsion allows microorganisms to explore their environments, from small bacteria \cite{braybook} to larger aquatic organisms \cite{stocker}. Since early  quantitative studies in the 1950s \cite{GITaylor,gray55},  theoretical and experimental studies of various microorganisms have led to discoveries of the numerous physical mechanisms used by motile cells to self-propel  through viscous fluids \cite{lighthill75,lauga_book}, in particular algae \cite{PedleyReview,goldstein14}, bacteria \cite{bacterial_hydrodynamics} and spermatozoa \cite{Sperm_review}.  The  quantitative investigation of such  phenomena has led to a better understanding of biological and physiological phenomena, including human reproduction \cite{fauci06} and  infectious diseases \cite{infections}. Beyond biology, the field of biomimetic design is showing a lot of promises in   mimicking these mechanisms to create controllable artificial microswimmers \cite{kim2017microbiorobotics,Hu201881,Dreyfus2005862,Williams2014,Huang2016}, which are now starting to find use in   medical applications \cite{Sitti2015205} such as targeted drug delivery \cite{Li2017} and minimally invasive surgeries \cite{Nelson201055}.
	
	One of the major practical design limitations for small-scale biomimetic  propulsion  was formulated by Purcell \cite{Purcell_scallop}.  It states that if an artificial swimmer undergoes periodic shape changes that are  ``reciprocal'' over a single period (i.e.~that remain identical under a time-reversible symmetry), the overall displacement of the swimmer is necessarily zero.
	The justification of this theorem lies in the time-reversibility of the Stokes equations that describe the motion of the surrounding fluid at the micro-scale.  As a result, one of the games to play for biomimetic  propulsion is to  circumvent the limitations of the scallop theorem \cite{life_around_scallop}.  The most obvious way is to use a propulsive mechanism that leads to non-reciprocal kinematics  \cite{nonreciprocal_kinematics,Tam2007,Avron2008,Najafi20044,Dreyfus2005161,Golestanian2010,Avron2005,Najafi2010,Iima2009} which in many cases is facilitated by the elastic compliance of the swimmer \cite{flexible_swimmer,Manghi2006,Qian2008,Cebers2005167}.  Other common ways include exploiting   inertia   \cite{childress_dudley_2004,scallop_inertia} or the non-Newtonian properties of the surrounding fluid \cite{Lauga_2009_EPL,Complex_fluids}.
	
	In this paper, we {investigate the motion of} artificial microswimmers based on a thermoresponsive hydrogel actuated under reciprocal cycles of heating and cooling. We first explore experimentally the propulsive features of these swimmers and find that the external actuation results in a slightly non-reciprocal shape change.
	Using standard  modelling of the hydrodynamic forces and flows for slender objects, we next show that the non-reciprocity in shape change is   not   sufficient to account for the overall swimming motion. Instead, we demonstrate that the propulsion  originates from the local flows induced by the  swelling and deswelling of the gel, with a mathematical model that leads to  excellent agreement with   experiments.

\begin{figure*}[t]
		\includegraphics[width=\linewidth]{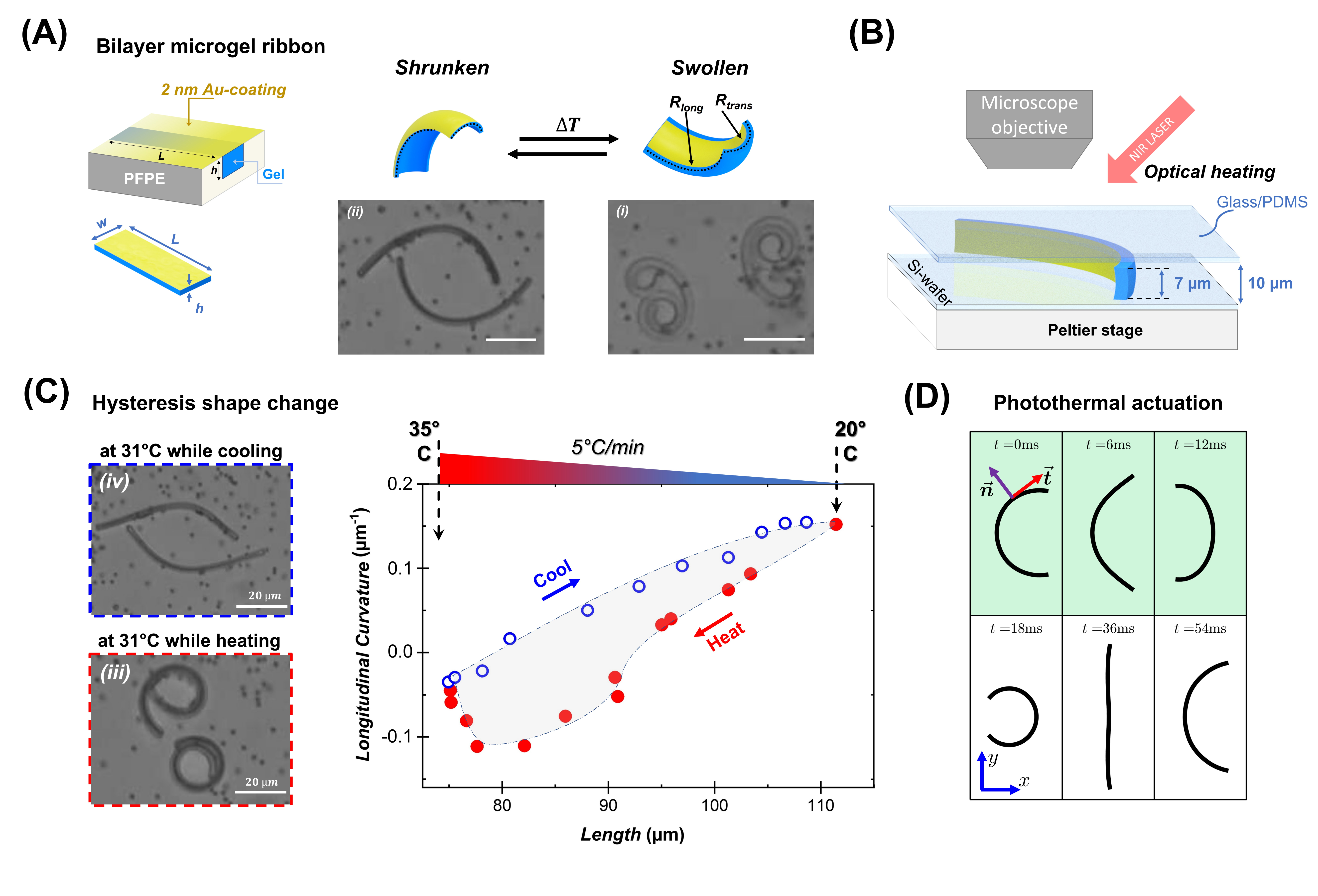}
		\caption{
			(A): A microgel ribbon made of crosslinked PNIPAm synthesized in a fluoroelastomer mould and sputtered with 2 nm thick gold as a passive layer. The ribbon has a length $L$ = 80~$\mu$m, a width $w$ = 5~$\mu$m and a thickness $t$ = 2~$\mu$m. When released into cold water, the gel swells and bends (with the gold layer inwards) along the longitudinal and transverse axis with a curvature $R_{long}$ and $R_{trans}$, respectively. Above 32$^\circ$C, the microgel shrinks and the curvatures reverse, with the gold layer outwards.
			(B): Experimental setup including a controlled heating stage, a microscope for optical observation and NIR laser irradiation at 45$^\circ$. The microgels are loaded into the microcompartment in which the ribbons spontaneously orient themselves edge-on relative to the confining surfaces, creating a gap of about 1.5~$\mu$m above and below the swollen gel. 
			(C): Variation of the longitudinal curvature (with $C_{long} = 1/R_{long}$) as a function of the ribbon length $L$ during a 5$^\circ$C/min heating (full, red symbols)/cooling (empty, blue symbols) ramp from 20$^\circ$C to 35$^\circ$C. Optical micrographs show the characteristic shapes of the ribbons swollen at 20$^\circ$C (i) and shrunken at 35$^\circ$C (ii). The graph shows the hysteresis of the ribbon curvature, along with the optical micrographs (iii) and (iv), taken at 31$^\circ$C, where the ribbon length is close to the preparation state ($L$ = 80~$\mu$m), but shows a significant difference in curvature depending on whether it is being cooled or heated.
			(D): Shape of the gel centreline during periodic photothermal actuation, extracted from the experimental videos and averaged over many periods, for $t_{\rm off} = 60~ms$, $t_{\rm on} = 12~ms$. Light green background indicates the part of the period when the microgel is heated (laser is on). }\label{fig::aachen}

\end{figure*}

	Microswimmers were produced as slender ribbons made of crosslinked poly(N-isopropyacrylamide)
	PNIPAm \cite{AA1,AA2,AA3,Deformable_helix,AA4a}, laden with gold nanorods and sputtered with a layer of gold on one of its largest sides, as illustrated in Fig.~\ref{fig::aachen}A. The microswimmers were placed in the  tight space between two flat rigid surfaces (Fig.~\ref{fig::aachen}B) so that the resulting motion can be considered to be quasi-two-dimensional (2D). For more information on the experimental methods, we refer the reader to the \hyperref[sec::supp_info]{supplementary material} (SM).

In order to actuate the microswimmers, we take advantage of the thermoresponsivity of the hydrogel. An increase in temperature reduces the solubility of the polymer, thereby shifting the equilibrium state, i.e.~the balance between the free energy of mixing the solvent molecules and polymer chains and the elastic free energy of the network \cite{AA5,AA6,MDoi,AA6a}. Due to the entropic nature of the hydrophobic interaction, water is transported out of the gel upon collapse, until a new balance between elastic and osmotic forces has been reached. In the case of a temperature-driven volume phase transition, the temperature change depends on   heat transfer, which is orders of magnitude faster than mass transport. Consequently, the imbalance between osmotic and elastic forces depends considerably on the rate of the temperature change and the dimensions of the microgel.

The key property that allows for the directional motion of the microgels is the anisometry of their bodies, which we control by using particle replication in non-wetting template \cite{AA7}. Such technique is valuable for the synthesis of a well-defined composition and shape. It allows the integration of plasmonic nanoparticles for optical heating and the possibility of a second passive layer. Such a bilayer bends with the temperature (see Fig.~\ref{fig::aachen}C), converting volume change into elastic energy \cite{AA8}. Under {near infrared laser (NIR)} irradiation, the gold nanorods absorb light energy and heat the polymer matrix, while the surrounding fluid acts as a heat sink \cite{AA9,AA10}. Rapid optical heating inevitably creates heterogeneities because the response of the polymer is slow, requiring diffusion of water molecules out of the network \cite{AA12}. The anisometry of the microgels and the transient heterogeneity cause stress accumulation resulting in body deformation. The process, although with hysteresis, is reversible and the shape recovers as soon as the irradiation ends \cite{AA13}.

\begin{figure*}
	\includegraphics[width=\linewidth]{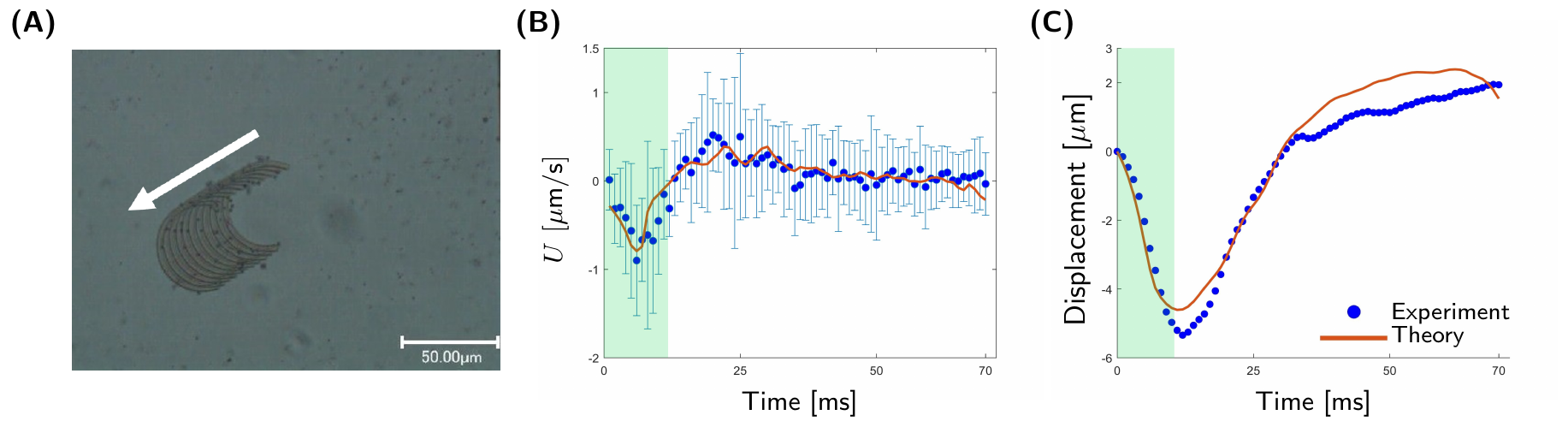}
	\caption{Experimental (symbols) and theoretical (lines) results for the translation of the microgel for irradiation times $t_{\rm off} = 60$~ms and  $t_{\rm on} = 10$~ms. Light green background indicates a part of the period when the laser is on. (A): Overlay of the gel shapes, sampled at the same phase of the periodic motion, with consecutive images being 4 periods apart. The arrow indicates the direction of motion. (B): Blue circles
		represent experimental results for the translational velocity; 
		error bars	 represent one standard deviation in estimating the mean velocities. The solid red line represents the  prediction from the theoretical model, Eq.~\eqref{eq::lmsmodel} with $\alpha=1.01$. (C): Displacement of the geometric centre-of-mass of the gel during a single period, as obtained by integrating the translational velocity in time. }\label{fig::results}
\end{figure*}

In our experiments, we induce a  periodic shape deformation of the microgel, of period $T = t_{\rm on}+t_{\rm off}$, by irradiating it steadily with a NIR light for a time $t_{\rm on}$ and letting it recover for a time $t_{\rm off}$. An example of the resulting shape deformation is shown in Fig.~\ref{fig::aachen}D, where we extracted 
the centreline of the microgel from experimental videos and averaged over a large number of periods (>100). For a range of irradiation and recovery times, we discover a non-vanishing net displacement of the gel over a single period of actuation. We illustrate these results in Fig.~\ref{fig::results} for $t_{\rm off} =$ 60~ms and $t_{\rm on} =$ 10~ms. In Fig.~\ref{fig::results}A we show a time lapse of the gel's motion, while we plot the average instantaneous speed $U$ of the gel in Fig.~\ref{fig::results}B and the average displacement of its geometric centre-of-mass in Fig.~\ref{fig::results}C.

It is clear from Fig.~\ref{fig::results}A that the gel translates in a direction that is not parallel to its apparent axis of symmetry. Given our setup, this might appear unexpected as the system is quasi-2D and the 2D shape of the gel appears to remain symmetric with respect to its initial axis of symmetry (i.e.~the $x$ axis in Fig.~\ref{fig::aachen}D). However, in our previous studies \cite{Deformable_helix,AA9} we observed that even the equilibrium shapes of such microgels are slightly chiral. We believe that the microgels investigated in this paper are also slightly chiral, around the axis perpendicular to the plane of observation, which causes the observed misalignment between the apparent axis of symmetry and the direction of translation. For a different choice of $t_{on}$, we observe a motion that is a combination of rotation and translation (see Fig.~S1A in the \hyperref[sec::supp_info]{SM}). Similarly, we believe that the chirality is responsible for this as well.  

Since the gel undergoes a non-reciprocal shape change (as illustrated in Fig.~\ref{fig::aachen}C), we might suspect that these shape changes, combined with the drag from the surrounding fluid, are sufficient to explain the observed net translation.  We tested this using  resistive-force theory of slender filaments \cite{Cox1970791} to compute the viscous drag on the gel as a function of its shape and  of the  ratio of drag coefficients $\Gamma = \xi_{\perp}/\xi_{\parallel} >1 $. In an infinite fluid $\Gamma\approx 2$;  due to the immediate presence of the walls, this drag coefficient ratio could be different from this bulk value, and could also be  time-dependent since the microgel changes shape throughout a  period of actuation. However, independently of our choice for $\Gamma$, we find that the predicted velocities so obtained are one order of magnitude smaller than what we see in the experiments. We therefore conclude that the net motion of the gel  is induced by a different physical mechanism.

We propose here that the swimming of the gel results instead from  the  flux of water that is emanating from the gel itself, due to its swelling/shrinking. In a tight confinement, which is the situation in our experiments, this creates strong local flows whose associated viscous stresses result in a net propulsive force acting on the gel, and hence to  locomotion.  To verify this hypothesis, we derive a  theoretical  model for these flows, which we show  is able to reproduce quantitatively the  translational motion of the gels. {Note  that   existing models have been developed  to explain the motion of passive, strongly confined microgels in external flows \cite{Lindner_brinkman,Lindner_asymmetric,Lindner_microchannel}. 
In our case the microgel   self-propels in a quiescent fluid  and  our model focuses instead on the propulsion mechanism.}

We assume that the gel is quasi-2D and that it remains symmetric with respect to the $x$ axis, a justified approach given the centreline shapes in Fig.~\ref{fig::aachen}D. In the centre-of-mass body frame, a material point $\bm{x}$  on the surface of the gel, $S$, is assumed to move (due to  shape changes)  with velocity $\bm{u}_S(\bm{x})$ while the surface of the gel  is ejecting fluid  with local velocity $\phi(x)\hat{\bm{n}}$, where  $\hat{\bm{n}}$ is the unit normal to $S$ at $\bm{x}$, pointing into the fluid. The aim of the model is to predict the translational velocity $U(t) \bm{e}_x$ of the gel's centre-of-mass. 

To proceed, we use  the theoretical framework developed for  jet propulsion in the absence of inertia \cite{jets_theory}. Following that work, we can compute the swimming speed of the gel by
applying  the  Lorentz reciprocal theorem of Stokes flows   using two flows: (i) the flow resulting from the deformation of the gel (i.e.~the current problem of interest) and (ii)  a second  `test' flow resulting from the translation of the gel with an arbitrary prescribed velocity $\tilde{U}\bm{e}_x$. Using $\tilde{\boldsymbol{\sigma}}$ to denote  the stress exerted on the gel in  the test flow leads to the integral identity \cite{jets_theory}
\begin{equation}
	\bm{0} = \int_S [U\bm{e}_x+\bm{u}_S+\phi(\bm{x})\hat{\bm{n}}]\cdot\tilde{\boldsymbol{\sigma}}\cdot \hat{\bm{n}}\,dS. \label{eq::jets}
\end{equation}
We  assume that the fluid ejection speed is uniform along the permeable sides of the microgel and, given that the flow is due to volume change in the gel, we may relate the rate of flow in and out of the gel to the rate of change of the gel length as
\begin{equation}
	\phi  = - \alpha \dot{L}, \label{eq::flux}
\end{equation}
where $\alpha$ is a dimensionless order-one parameter. 

The relationship in Eq.~\eqref{eq::flux} assumes implicitly that the rate of change of microgel's length  $L$, is the only input in the model on the fluid transport and is therefore sufficient to capture the asymmetry between the swelling and the shrinking process, experimentally demonstrated, for example, as the hysteresis in Fig.~\ref{fig::aachen}C. 
In an idealised case, $\alpha$ would represent the ratio between the full cross-sectional area of the microgel and the portion of area that is permeable. However, as the geometry in our experiments  is  rather intricate, we let $\alpha$ be the only fitting parameter of the model.

The result in Eq.~\eqref{eq::jets}, which is a surface integral, can  be next coarse-grained over the thin cross-section of the gel and  turned into a line integral along the length $L$ of the gel cross-section. To approximate the average of the stress $\tilde{\boldsymbol{\sigma}}\cdot \hat{\bm{n}}$ due to the test flow acting on a cross-section,  we use the resistive-force theory of slender filaments \cite{Cox1970791}, stating that $\tilde{\boldsymbol{\sigma}}\cdot \hat{\bm{n}}\propto \left(\bm{t}\bm{t}+\Gamma \bm{n}\bm{n}\right)\cdot\tilde{U}\bm{e}_x$. With the choice $\Gamma=2$ (wall and unsteady effects were  found to be subdominant  in our comparison with the experiments), we then obtain the integral equation
\begin{eqnarray}\label{eq:2}
	0 = \int_L [U\bm{e}_x+\bm{u}_S-\alpha \dot{L} \bm{n}]\cdot[\bm{t}\bm{t}+2 \bm{n}\bm{n}]\cdot \bm{e}_x \,dl,
\end{eqnarray}
where $\bm{n}$ is now the unit normal pointing into the fluid but on the non-plated side of the gel, where most fluid ejection occurs, and $\bm{t}$ is the unit tangent along the centreline (see Fig.~\ref{fig::aachen}D). The integral is performed along the centreline of the gel with $dl$ being its line element. We can now use Eq.~\eqref{eq:2} to derive the instantaneous translational velocity as
\begin{equation}
	U = \frac{2\alpha \dot{L}\int_L n_x \, dl-\int_L \bm{u}_S\cdot[\bm{t}\bm{t}+2 \bm{n}\bm{n}]\cdot \bm{e}_x \,dl}{\int_L (t_x^2+2 n_x^2)\,dl}. \label{eq::vel_model}
\end{equation}

{We note that our  model does implicitly  include the  strong hydrodynamic confinement of the gel. The normal stresses and therefore the perpendicular drag force acting on a cross-section are dominated by the pressure drop    required to force the fluid through the narrow gap. This suggests that the integral of $\phi\hat{\bm{n}}\cdot\tilde{\boldsymbol{\sigma}}\cdot\hat{\bm{n}}$ is coarse-grained to a line integral with the integrand proportional to $\bm{n}\cdot\bm{e}_x$, as done above. The confined geometrical setup is therefore implicitly included in our  model.}

To test our theoretical prediction, we use the kinematics for the  gel's centreline extracted from experimental videos. This was done using the finite difference method on the  centreline shapes. 
The videos were recorded at frame rate $\Delta t=1$~ms with actuation periods $T=N\Delta t$ that were always an integer multiple of $\Delta t$. This allowed us to track the gel over many periods so that, within each period, we record the shape at exactly the same set of phases $t_n = n \Delta t$, with $n = 0,1,\dots,N$. Thus, by averaging the extracted shapes and gel displacements over many periods of actuation ($\approx$ 100), we greatly improve the accuracy of our finite difference approach. 

The theoretical prediction for $U_n = U(n\Delta t)$  in Eq.~\eqref{eq::vel_model} can then be written as
\begin{equation}
	U_n = \alpha F_n +U_n^{S}, \label{eq::lmsmodel}
\end{equation}
where
\begin{equation}
	F_n = \frac{2 \dot{L}\int_L n_x \, dl}{\int_L (t_x^2+2 n_x^2)\,dl}, 
	\quad	U_n^{S} = -\frac{\int_L \bm{u}_S\cdot[\bm{t}\bm{t}+2 \bm{n}\bm{n}]\cdot \bm{e}_x \,dl}{\int_L (t_x^2+2 n_x^2)\,dl}, 
\end{equation}
with $F_n$ representing the contribution due to the fluid transport through the surface of the gel and $U_n^{S}$ that due to the evolution of the shape. Note that taking $F_n=0$ is equivalent to assuming the locomotion is due solely to the non-reciprocal shape changes, which is not able to reproduce our experimental results (see above). 

The only free   parameter in the model is the {order-one} dimensionless constant of proportionality $\alpha$ from Eq.~\eqref{eq::flux}. We fix its value by performing  a least-square fitting on the model given by   Eq.~\eqref{eq::lmsmodel} where $U_n$ is extracted from the experiments while $F_n$ and $U_n^{S}$ are computed   using the kinematics extracted from the experimental data, as described above. The optimal value of $\alpha$ was found to be slightly different for different experimental conditions which is not surprising given that it encapsulates complicated and highly sensitive microgel swelling/shrinking dynamics.

In   Fig.~\ref{fig::results}B-C we show the comparison between the experimental data (blue symbols) and the theoretical model (solid red line). Our theoretical prediction for the translational velocity is in excellent agreement with the experimental results for both the instantaneous value of the velocity (Fig.~\ref{fig::results}B) and the total integrated displacements (Fig.~\ref{fig::results}C). Moreover, we obtain the same level of agreement in an experiment with a different set of irradiation times that lead to a significant rotation of the microgel (see Fig.~S1 in the \hyperref[sec::supp_info]{SM}).

\begin{figure}[t]
	\includegraphics[width=\linewidth]{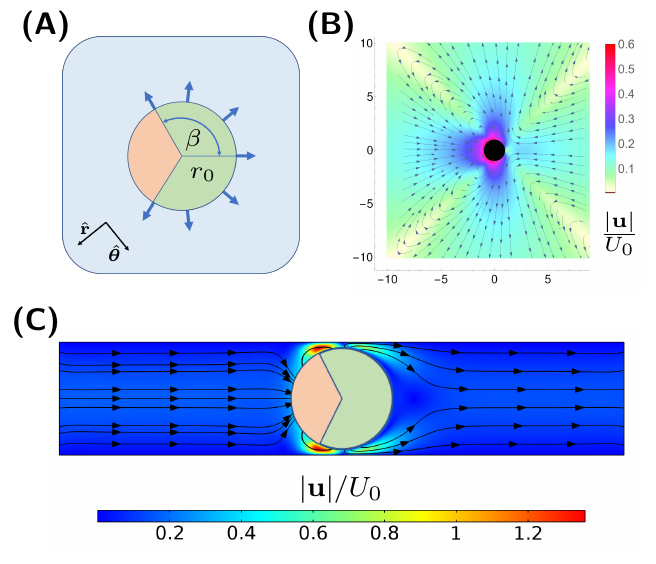}
	\caption{Using a two-dimensional model where the gel expels  flow uniformly along its surface (A), we show the  flow induced around the circular cross-section in an unbounded fluid [(B), analytical solution] and in the tight space between two rigid surfaces [(C), numerical computations]. (A): Sketch of a toy model of a cross-section with surface angle $2\beta < 2\pi$ expelling fluid. (B): Flow around the model cross-section in an unbounded domain, given by Eqs.~\eqref{flow_u}-\eqref{flow_v} for $\beta=2\pi/3$. Arrows represent the flow direction while colours indicate the flow magnitude non-dimensionalised by $U_0$. (C): Results of COMSOL \cite{comsol} simulations for a model cross-section ($\beta = 2\pi/3$) in presence of two no-slip plane walls (top and bottom). Lines show streamlines while  colours indicate the flow magnitude non-dimensionalised by $U_0$ (spacing between the walls of $2r_0/0.9$).}\label{fig::flow}
\end{figure}

We further investigate the hydrodynamics of  swimming  by predicting theoretically the fluid flow around a cross-section of the microgel. As a model system,  we assume that the  cross-section of the gel is circular (`disk' in what follows) of radius $r_0$ and is ejecting (taking in) fluid at speed $U_0$ from the part of its perimeter that describes an angle $2\beta < 2\pi$ (see sketch in Fig.~\ref{fig::flow}A). As the fluid is ejected (taken in) the radius of the gel must change as $\dot{r}_0 = -\beta U_0 /\pi$ due to the conservation of volume.
In an unbounded fluid,  an analytic solution exists for the   flow around this two-dimensional setup~\cite{blake_flow}. Under the assumption of zero  net hydrodynamic force, the disk translates with velocity $-U_0\sin \beta / \pi$ and drives a two dimensional flow  $u\hat{\bm{r}} + v\hat{\bm{\theta}}$  in  polar coordinates given by
\begin{eqnarray}
	\label{flow_u} \frac{u}{U_0} &=& \frac{\sin \beta \cos \theta}{\pi }\left(\frac{r_0}{r}\right)^2  \\ 
	&&+\sum_{n=2}^{\infty} \frac{\sin n\beta \cos n\theta}{\pi n} \left[n\left(\frac{r_0}{r}\right)^{n-1}-(n-2)\left(\frac{r_0}{r}\right)^{n+1} \right] \hspace{2mm}, \nonumber \\
	\label{flow_v} \frac{v}{U_0} &=& \frac{\sin \beta \sin \theta}{\pi}\left(\frac{r_0}{r}\right)^2  \\ 
	&&+\sum_{n=2}^{\infty} \frac{n-2}{\pi n}\sin n\beta \sin n\theta \left[\left(\frac{r_0}{r}\right)^{n-1}-\left(\frac{r_0}{r}\right)^{n+1} \right] \nonumber .
\end{eqnarray}
We illustrate this unbounded flow (non-dimensionalised by $U_0$) in  Fig.~\ref{fig::flow}B in the case $\beta = 2\pi/3$ (the flow shown is the series  truncated at order $(r_0/r)^{n}$, with $n=40$). 

If we now confine the disk between rigid  surfaces,  the strongest flow occurs in the narrow gaps between the disk and the walls, and the analytical solution that could capture this part of the flow would be cumbersome. To gain intuition about flow features in this case, we use COMSOL \cite{comsol} to solve for the Stokes flow under the same situation  (i.e.~force-free  locomotion of the disk with $\beta = 2\pi/3$) for a disk symmetrically located in a space of size $2r_0/0.9$ between two rigid surfaces.
We show in  Fig.~\ref{fig::flow}C the  streamlines and the magnitude of the flow,    illustrating the strong flows that are locally induced in the tight space between the disk and the surfaces and their right-left asymmetry -- this is the  physical origin of the  net propulsion.

In summary, in this paper we report a {jet-driven} type of microgel propulsion. In contrast with many studies of artificial    swimming, which rely on the non-reciprocal bending of  the centerline of slender swimmers to swim,  here locomotion is 
driven by the localised  fluid transport in and out of the gel caused by its rapid swelling and deswelling. 
This viscous-jet hypothesis is confirmed to be  the dominant mechanism of propulsion by a  hydrodynamic model that shows excellent agreement with our experimental results. 

\section*{Supplementary information}
Details of the experimental methods.

\section*{Acknowledgements}

This project has received funding from the European Research Council under the European Union's Horizon 2020 Research and Innovation Programme (Grant No. 682754 to E.L.) and from Trinity College, Cambridge (IGS scholarship to I.T.). This research was supported by the German science foundation, through the priority program on microswimmers Grant No. 255087333. We have also benefited from the thriving scientific environment provided by the SFB 985 on functional microgels and microgel systems.

\section*{Data availability}
The experimental data is available on request from the corresponding author. 

%

\newpage

\appendix

\section*{Supplementary material \label{sec::supp_info}}

\begin{figure*}
	\includegraphics[width=\linewidth]{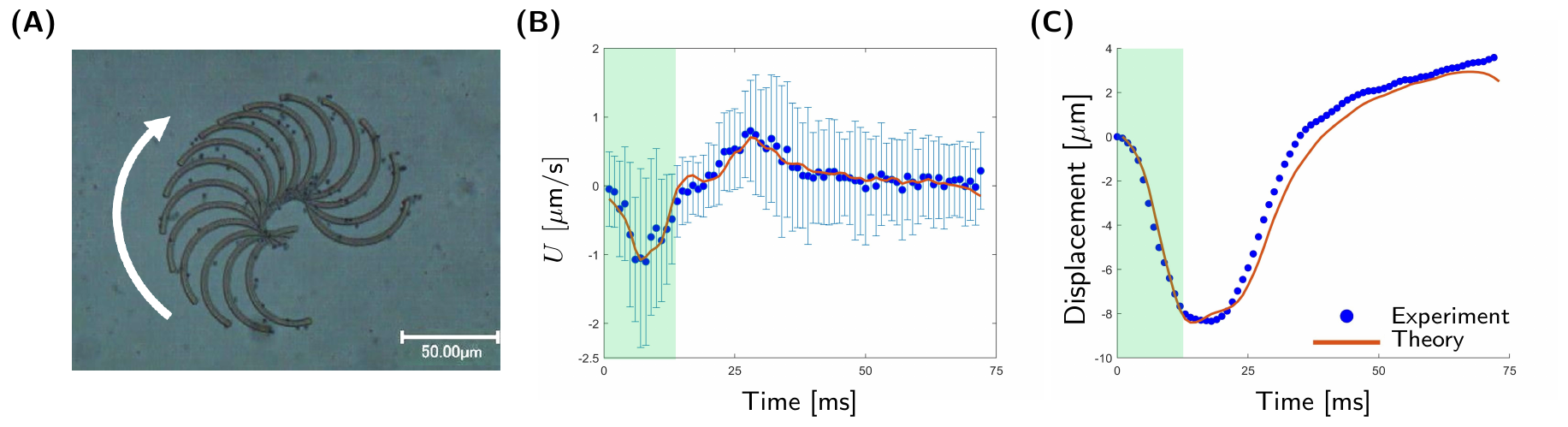}
	\caption{Experimental (symbols) and theoretical (lines) results for the translation of the microgel for irradiation times $t_{\rm off} = 60$~ms and  $t_{\rm on} = 12$~ms. Light green background indicates a part of the period when the laser is on. (A): Overlay of the gel shapes, sampled at the same phase of the periodic motion, with consecutive images being 4 periods apart. The arrow indicates the direction of motion. (B): Blue circles
		represent experimental results for the translational velocity; 
		error bars	 represent one standard deviation in estimating the mean velocities. The solid red line represents the  prediction from the theoretical model, Eq.~(5) with $\alpha=1.25$. (C): Displacement of the geometric centre-of-mass of the gel during a single period, as obtained by integrating the translational velocity in time. }\label{fig::results_supp}
\end{figure*}

\subsection*{Soft template moulding of bilayer hydrogel ribbons
}

The microswimmers were prepared by particle replication in nonwetting templates (short PRINT \cite{AA7}), which has been modified to yield bilayer structures as described in our previous reports \cite{AA11}. Preparation of the mould: A microstructured, (25 x 50) mm Si-Wafer with an array of rectangular bars of L = 80~$\mu$m, w = 5~$\mu$m and h = 2~$\mu$m (AMO GmbH) was replicated using a fluorinated elastomer (PFPE, Fluorolink MD 40, Solvay Solexis). Therefore, a liquid PFPE containing 1 wt\% of photoinitiator 2-Hydroxy-2-methylpropiophenone (Sigma-Aldrich) was prepared an degassed by bubbling with nitrogen. A (25 x 50 mm) glass slide was placed on the wafer using double layer of Parafilm strips as spacer (total thickness: 200~$\mu$m, Bemis). This gap was therefore filled with PFPE liquid and subsequently cured with UV irradiation (370 nm, 2 W) for 10 min. This results an elastomeric film that is easy to separate from the template, which we wash with acetone to remove any unreacted substances. 
\subsection*{Preparation of the bilayer hydrogels
}

The hydrogel was prepared from a UV curable solution containing 575 mg N-Isopropylacrylamid (NIPAm, 97\%, recrystallized in n-hexane), 7.83 mg crosslinker N,N’-Methylenebisacryamide (BIS, 99\%), 11.4 mg photoinitiator 4’-(2-hydroxyethoxy)-2-methylpropiophenone (98\%) and 981~$\mu$L DMSO as solvent, containing the PEGylated gold nanorods. The concentration of nanorods was adjusted to an optical density of 240 at the maximum absorption (typically around 810 nm), which was measured by UV-VIS spectroscopy. The gelation was performed in a homemade chamber including a quartz glass window to apply a pressure of approximately 260 kPa, while maintaining UV transparency for curing. Typically, a (6 x 6) mm piece of the mould was placed on the base of the chamber and 0.5~$\mu$L of the prepolymer solution were applied. A flat (5 x 5) mm featureless PFPE film (thickness: 400~$\mu$m, produced similarly to the description above) was placed on the mould and press-fitted through the quartz window. After UV curing for 20 min (366 \& 254 nm, 8W), the mould was separated and a 2 nm gold layer was applied by sputter-coating (60 mA, 10 s, Edwards S150B). A (25 x 25) mm glass slide was plasma treated (O$_2$, 40 mL/min, 200 W, 5 min, PVA TePla 100) and a droplet of 714 g/L solution of Polyvinylpyrrolidone (PVP, 40 kDa, Sigma-Aldrich; in ethanol/water, volume ratio: 7/63) was applied, into which the sample was placed upside down. The sample was then left to dry, allowing the hydrogel bind temporarily to the glass slide and thus peel it off the mould.

\subsection*{Production of chambers for microscopic investigations
}

PDMS-moulded chambers with a height of 10~$\mu$m and a footprint of (600 x 600)~$\mu$m were produced to confine the movement of the swimmers. The master structures were designed using a CAD program (Inventor, Autodesk), exported as .stl file and transferred to a multiphoton lithography setup (Photonic Professional GT, Nanoscribe GmbH) to 3D print the structures onto a glass slide. Here, a 25x immersion objective (numerical aperture (NA), 0.8; Zeiss) and IP-S photoresin (Nanoscribe GmbH) were used. The slicing and hatching distance were 1~$\mu$m and 0.5~$\mu$m, respectively. The laser power was set to 100\% and scanning speed at 100 mm/s. Prior to use, glass slides were plasma treated (O$_2$, 40 mL/min, 200W, 5 min, PVA TePla 100) and incubated overnight in a solution containing 1\% 3‑(Trimethoxysilyl)propyl acrylate (Sigma-Aldrich) in acetone. This allowed covalent binding of the printed material to the glass substrate. After print, excess resin was washed by incubation in 1-methoxy-2-propanol acetate (Sigma-Aldrich) for 10 min and in isopropanole (Sigma-Aldrich) for 3 min. The obtained master was then replicated using PDMS. Typically, 0.2 mL of a degassed PDMS (Sylgard 184, Dow Corning) was added and covered with a glass slide. Square capillaries (Vitrocom) with a width of 600~$\mu$m were used as spacers, to give a total film thickness of 600~$\mu$m. The PDMS was cured at 90$^\circ$C for 60 min, separated from the master and washed with isopropanol.

\subsection*{Filling of the chambers with microswimmers
}
A Si-Wafer was diced to (20 x 20) mm, cleaned with ultrapure water (0.055~$\mu$S/cm, Purelab Plus UV) and plasma treated (O$_2$, 40 mL/min, 200W, 5 min, PVA TePla 100). To release the hydrogels from the mould, the latter was peeled and subsequently 2-3~$\mu$L of ultrapure water (0.055~$\mu$S/cm, Purelab Plus UV) were added by pipetting. After sufficient time for the swimmers to release, 2~$\mu$L of this dispersion were pipetted onto the freshly prepared Si-Wafer and dried at 35$^\circ$C. The moulded PDMS was cut to (15 x 15) mm and a (25 x 25) mm glass slide (thickness 150~$\mu$m) was cleaned with ultrapure water. Both were then treated with plasma (O$_2$, 30 mL/min, 100W, 30 s, PVA TePla 100). This step increases the hydrophilicity of the PDMS to enhance filling with water and avoid adhesion of the hydrogel, while the 150~$\mu$m thin glass slide improves the mechanical stability of the setup. Depending on the objective of the experiment either 2~$\mu$L of ultrapure water, or 2~$\mu$L of a microsphere dispersion (FluoSpheres$^\text{\textregistered}$, 1.0~$\mu$m, Invitrogen; 1:10 volume ratio in ultrapure water) was pipetted onto the wafer. Immediately, the PDMS film was used to trap the swimmers in the chambers and the glass slide was added to gently apply pressure and remove excess liquid. The sample was kept at 40$^\circ$C for a few minutes to evaporate excess liquid and ensure sealing.

\subsection*{NIR-light actuation and microscopic investigation}The hydrogels were actuated by pulse sequences of near infrared LASER irradiation (808~nm, 10 W, Roithner Lasertechnik) at an incident angle of 45$^\circ$, giving an elliptical spot size of approximately (2~x~1.5)~mm. The temperature of the sample was kept at (20 $\pm$~0.1)$^\circ$C, using a water cooled peltier element mounted onto the microscope stage. Recording videos with high frame rates (1~frame/ms) was realized using a digital microscope (VW-9000D, Keyence) with a high-speed camera.

\end{document}